\documentclass[pra,aps,twocolumn,preprintnumbers,showpacs,superscriptaddress]{revtex4}

\usepackage{latexsym,epsfig,amssymb,amsfonts,amsmath,graphicx}

\newcommand{\ket}[1]{\left | \, #1 \right \rangle}
\newcommand{\bra}[1]{\left \langle #1 \, \right |}

\newcommand{\tr}[1]{\mbox{Tr} \, #1 }

\newcommand{\equref}[1]{Eq.~(\ref{#1})}
\newcommand{\equsref}[2]{Eqs.~(\ref{#1}, \ref{#2})}
\newcommand{\figref}[1]{Fig.~\ref{#1}}

\newcommand{\pur}[2]{\mbox{Tr}\,\left\{#1 ^2_{#2}\right\}}
\newcommand{\avpur}[2]{\overline{\mbox{Tr}\,\left\{#1 ^2_{\left(#2\right)}\right\}}}

\def\id{\mathbb{I}}

\allowdisplaybreaks

\begin{document}

\title{Detection and characterization of multipartite entanglement in optical lattices}

\author{R.N. \surname{Palmer}}%
\author{C. \surname{Moura Alves}}%
\author{D. \surname{Jaksch}}%
\affiliation{Clarendon Laboratory, University of Oxford, Parks Road, Oxford OX1 3PU, UK.}%

\date{\today}

\begin{abstract}
We investigate the detection and characterization of entanglement
based on the quantum network introduced in [Phys.~Rev.~Lett.~{\bf
93}, 110501 (2004)] for different experimental scenarios. We first
give a detailed discussion of the ideal scheme where no errors are
present and full spatial resolution is available. Then we analyze
the implementation of the network in an optical lattice. We find
that even without any spatial resolution entanglement can be
detected and characterized in various kinds of states including
cluster states and macroscopic superposition states. We also study
the effects of detection errors and imperfect dynamics on the
detection network. For our scheme to be practical these errors
have to be on the order of one over the number of investigated
lattice sites. Finally, we consider the case of limited spatial
resolution and conclude that significant improvement in
entanglement detection and characterization compared to having no
spatial resolution is only possible if single lattice sites can be
resolved.
\end{abstract}


\maketitle

\section{Introduction}

Multipartite entanglement is an essential ingredient for complex
quantum information processing (QIP) tasks, such as quantum
error-correction \cite{QEC}, multi-party quantum communication
\cite{GHZ} and universal quantum computation in the one-way model
\cite{Cluster}. It has been generated between photons
\cite{4photon} and in a controlled way both in ion traps
\cite{IonTraps} and in experiments with Mott insulating states of
neutral bosonic atoms in optical lattices \cite{BlochEnt,BECinOL}.
For photon and ion trap experiments full state tomography
\cite{4photon,MLEtomography} impressively proved the existence of
multipartite entanglement with few particles. However, its
unambiguous detection in optical lattice setups has so far not
been possible. The measurements implemented in the actual
experiments only provided information on the average density
operator of each individual atom (representing a qubit), but not
on the density operator of the composite system \cite{BlochEnt}.
Full state tomography in optical lattices seems to be infeasible
due to the large number of involved atoms and thus other methods
are required to prove the existence of entanglement. Recently, a
quantum network that can detect multipartite entanglement in
bosons, and can be realized in optical lattices, was proposed
\cite{Carolina2004}.

This quantum network is made up of pairwise beam-splitters (BS)
acting on two identical copies of a multipartite entangled state
$\rho_{12\dots n}$ of $n$ atoms, and allows the determination of
the purity of $\rho_{12\dots n}$ and of all its reduced
subsystems. The purities provide us with a separability criterion
based on entropic inequalities \cite{Carolina2004} that, if
violated, indicate that the state $\rho_{12\dots n}$ is entangled.
This nonlinear entanglement test can be implemented experimentally
with a number of copies of $\rho_{12\dots n}$ independent of $n$,
rather than the exponentially large number of copies needed for a
full state tomography, and is more powerful than the other common
experimental methods, such as Bell inequalities \cite{Bell64} and
entanglement witnesses \cite{Witness}.  In fact, the nonlinear
inequalities are known to be strictly stronger than the Bell-CHSH
inequalities \cite{Entropies}. Furthermore, if it may be assumed
that the entangled state is characterized by a few parameters only
it is often possible to determine them from the results obtained
from the entanglement detection network.

The main aim of this paper is to analyze the operation of the
entanglement detection network under realistic experimental
conditions. After introducing the ideal scheme we consider the
particular realization of the network in an optical lattice. We
discuss three main sources of imperfections: limited or no spatial
resolution, errors in the BS operations occurring with a
probability $q$, and failure to detect an atom with probability
$p$. In current experiments none of these errors can be suppressed
completely and thus it is important to find out whether their
presence still allows the unambiguous detection and
characterization of multipartite entangled states. Our main
results will be that (i) even without spatial resolution it is
possible to unambiguously detect and characterize entanglement in
multipartite states like cluster states and macroscopic
superposition states, (ii) the effects of BS and detection errors
can be made arbitrarily small by increasing the number of runs of
the entanglement detection network, and (iii) the required number
of experimental runs for achieving this is reasonably small if $k
q \lesssim 1$ and $k p \lesssim 1$, where $k\leq n$ is the number
of particles in the subsystem whose reduced purity is being
measured. Finally we will find that (iv) single site resolution is
necessary to achieve significant improvement compared to having no
spatial resolution at all.

The paper is organized as follows: in Sec.~\ref{secentdect} we
introduce the entropic inequalities and discuss how multipartite
entanglement can be detected and characterized by them. Then we
describe the entanglement detection network and show how it can be
realized in optical lattices and how the main sources of
imperfections arise. In Sec.~\ref{secentdectnospat} we investigate
the detection and characterization of multipartite entangled
states if no spatial resolution is available, and in
Sec.~\ref{sec:errors} we analyze the effect of the dominant
experimental errors. We also discuss the case of limited spatial
resolution in the measurements. Finally we summarize our results
in Sec.~\ref{secconcl}.

\section{Entanglement Detection and Characterization in Optical Lattices}
\label{secentdect}

In this section we first present the entropic inequalities which
are used to detect multipartite entanglement. We show how they can
be utilized to characterize multipartite entanglement if the
entangled state is described by few parameters only. Then we
introduce our entanglement detection network and discuss its
implementation in optical lattices. Finally we study the main
types of imperfections and resulting errors that we expect to be
present in this experimental realization.

\subsection{Entropic inequalities}

Our network uses the entropic inequalities introduced
in~\cite{Carolina2004} for multipartite entanglement detection.
These inequalities provide a set of necessary conditions for
separability in multipartite states. Consider a state
$\rho_{123\dots n}$ of $n$ subsystems. If $\rho_{123\dots n}$ is
separable, then
\begin{eqnarray}
\rho_{123\dots n}= \sum_{\ell} C_\ell \rho^{\ell}_1 \otimes
\rho^{\ell}_2 \otimes \rho^{\ell}_3 \otimes \ldots \otimes
\rho^{\ell}_n, \label{rho}
\end{eqnarray}
where $\rho^{\ell}_j$ is a state of subsystem $j$, and
$\sum_{\ell} C_\ell=1$. The purity $\pur{\rho}{B}$ of the reduced
density operator $\rho_{B}$ with $B\subseteq\{1,2,\dots,n\}$ is
less than or equal to the purity of any of its reduced density
operators:
\begin{equation}
\pur{\rho}{A}\geq\pur{\rho}{B}\text{ for all }A\subseteq B.
\label{ineq}
\end{equation}
We can use these inequalities to give a relation between the
average purities of all reduced density operators of a given
number $k$ of subsystems defined by
\begin{equation}
\avpur{\rho}{k}=\left[{n\choose
k}\right]^{-1}\sum\limits_{|B|=k}\pur{\rho}{B},
\end{equation}
where $B$ is summed over all combinations of $k$ subsystems. For
completeness we define $\avpur{\rho}{0}\equiv 1$. From
\equref{ineq} we find that for any separable state
\begin{equation}
\avpur{\rho}{k} \geq \avpur{\rho}{k'} \text{ for all } k \leq k'.
\label{ineq2}
\end{equation}
We will now consider how \equsref{ineq}{ineq2} can be used to
detect entanglement and to characterize entangled states.

\subsubsection{Entanglement detection}

Any state $\rho$ that violates any of the inequalities
\equsref{ineq}{ineq2} is entangled. If $\rho_{123\dots n}$ is
separable and pure, $\pur{\rho}{123\dots n}=1$ and all the above
inequalities become equalities; since a state with all
$\pur{\rho}{B}=1$ is necessarily a product state, all pure
entangled states violate \equsref{ineq}{ineq2}. All pure entangled
states can hence be detected by comparing $\pur{\rho}{123\dots n}$
with any other $\pur{\rho}{B}$ or $\avpur{\rho}{k}$ unlike
entanglement witnesses or Bell inequalities where different states
often require different witnesses. From the Schmidt decomposition
of any pure state into two disjunct subsystems $A$ and $B$ with $A
\cup B = \{1,2\dots n\}$ we find $\pur{\rho}{A} = \pur{\rho}{B}$
and hence $\avpur{\rho}{k} =\avpur{\rho}{n-k}$ for all pure
$\rho$.

Mixed entangled states do not always violate
\equsref{ineq}{ineq2}, but since $\pur{\rho}{B}$ is continuous in
$\rho$ a sufficiently small amount of noise added to a pure
entangled state will leave it still violating the inequalities. In
the examples studied in this paper the noise level at which the
inequalities no longer detect entanglement is a large fraction of
the level at which entanglement ceases to be present. The
permissible range values for the purities of any state $\rho$ is
\begin{equation}
2^{-k}\leq\pur{\rho}{B}\leq 1 \quad \text{or} \quad
2^{-k}\leq\avpur{\rho}{k}\leq 1, \label{purbounds}
\end{equation}
where $k=|B|$. The minimum is attained by the maximally mixed
state $\rho=2^{-n}\id$ (with $\id$ the identity operator) and the
maximum by any pure product state.

Our test can often verify that a state is truly $n$-particle
entangled and not just a collection of small entangled subsystems,
because $\pur{\rho}{B}<\pur{\rho}{}$ always indicates that $B$ is
entangled with the rest of the system and cannot be caused by
entanglement within $B$ itself. Furthermore our test is
insensitive to local unitary operations, which cannot alter the
entanglement structure, since they do not change the purities. It
is, however, in most cases sensitive to local measurements
destroying entanglement as these generally do change the purities.
If the tested state $\rho$ is of a known form and characterized by
few parameters only the kind and degree of violation of
\equsref{ineq}{ineq2} can be used to determine these parameters
and thus characterize the state.

\subsubsection{Characterization of entanglement: Macroscopic superposition states}

Macroscopic superposition states $\ket{\gamma_n}$ defined by
\begin{equation}
\ket{\gamma_n}=\frac{\ket{0}^n+(\gamma\ket{0}+\sqrt{1-|\gamma|^2}\ket{1})^n}{\sqrt{2+\gamma^n+\bar\gamma^n}},
\end{equation}
with a single complex parameter $\gamma$ satisfying $|\gamma|\leq
1$ arise naturally in several systems such as BECs \cite{cat-bec}
and superconductors \cite{cat-squid}. The quantity $n(1-
|\gamma|^2)$ has been suggested as a measure of the effective size
of the state, as in some respects $\ket{\gamma_n}$ is equivalent
to a maximally entangled state of $n(1-|\gamma|^2)$ particles
\cite{Cat}. The purities of $\ket{\gamma_n}$ are given by
\begin{eqnarray}
\pur{\rho}{B} &=&
\frac{2+2\gamma^k\bar\gamma^k+2\gamma^{n-k}\bar\gamma^{n-k}}{(2+\gamma^n+
\bar\gamma^n)^2} \nonumber + \\ && \frac{4\gamma^n +
4\bar\gamma^n+\gamma^{2n}+\bar\gamma^{2n}} {(2+\gamma^n+
\bar\gamma^n)^2}, \label{macropur}
\end{eqnarray}
where $k=|B|$. These purities violate \equref{ineq}, showing that
$\ket{\gamma_n}$ is entangled, for all $|\gamma|<1$ and measuring
the purities allows determining $\gamma$. For $\gamma=0$ the state
$\ket{\gamma_n}$ is a maximally entangled state which is pure
($\pur{\rho}{}=1$) while all its subsets have purity
$\pur{\rho}{B}=1/2$. Up to local unitaries it is the only state
with these purities, so can be unequivocally identified by our
test.

We demonstrate the effect of noise on our method by studying
individual dephasing of each qubit. Such dephasing noise maps a
state according to
\begin{equation}
\rho \rightarrow \sum_A
\left(1-\frac{d}{2}\right)^{n-|A|}\left(\frac{d}{2}\right)^{|A|}
\left(\prod_{j\in A}\sigma_{z,j}\right)\rho\left(\prod_{j\in
A}\sigma_{z,j}\right), \label{phasenoise}
\end{equation}
where $\sigma_{z,j}$ is a phase flip applied to qubit $j$, $A$ is
the set of qubits which undergo a phase flip and $0\leq d\leq 1$
is the decoherence parameter. Applying the map \equref{phasenoise}
to $\ket{(\gamma=0)_n}$ we find for the resulting density operator
\begin{eqnarray}
\rho&=&\frac{1}{2}\left(\ket{0}^n\bra{0}^n+\ket{1}^n\bra{1}^n\right) \nonumber \\
&&\qquad+\frac{(1-d)^n}{2}\left(\ket{0}^n\bra{1}^n+\ket{1}^n\bra{0}^n\right),
\end{eqnarray}
and thus $\pur{\rho}{}=(1+(1-d)^{2n})/2$ while $\pur{\rho}{B}=1/2$
for all subsets $B$. Therefore entanglement is detected whenever
it is present. This is not generally the case as can easily be
seen by looking at a Werner state $\rho =
(1-d)\ket{(\gamma=0)_n}\bra{(\gamma=0)_n} + 2^{-n} d \; \id$ for
which entanglement is detected if $d<1-(2^{n-1}+1)^{-1/2}$.
However, e.g.~in the case $n=2$ this state is entangled iff
$d<2/3$ \cite{Werner}, while our test works only for
$d<1-1/\sqrt{3}$.

\subsubsection{Characterization of entanglement: Cluster states}
\label{subsec:CharId}

Cluster and graph states are multipartite entangled states which
form the basic building blocks of the one-way quantum computer
\cite{Cluster}. We consider cluster-like states $\ket{\phi_n}$
defined by
\begin{equation}
\ket{\phi_n}=\frac{1}{\sqrt{2^n}}\prod_{j=1}^n(\ket{0}_j
e^{i\phi\sigma_{z,j+1}}+\ket{1}_j)=\frac{1}{\sqrt{2^n}}
\sum\limits_{x=0}^{2^n-1}e^{i\phi c(x)}\ket{x},
\end{equation}
where $c(x)$ is the number of occurrences of the sequence $01$ in
the $n$-bit binary number $x$. These states have already been
created in optical lattices \cite{BlochEnt} and represent a
cluster state for $\phi = \pi$. Current methods for detecting them
essentially perform a tomographic measurement of the average
single particle density matrix, which goes from pure
$(\ket{0}+\ket{1})(\bra{0}+\bra{1})/2$ at $\phi=0$ to maximally
mixed $\id/2$ at $\phi=\pi$ and back again at $\phi=2\pi$. This
method thus yields one measurement $\avpur{\rho}{1}$ relating to
entanglement and two measurements relating to local unitaries.

\begin{figure}
\includegraphics[width=8.5cm]{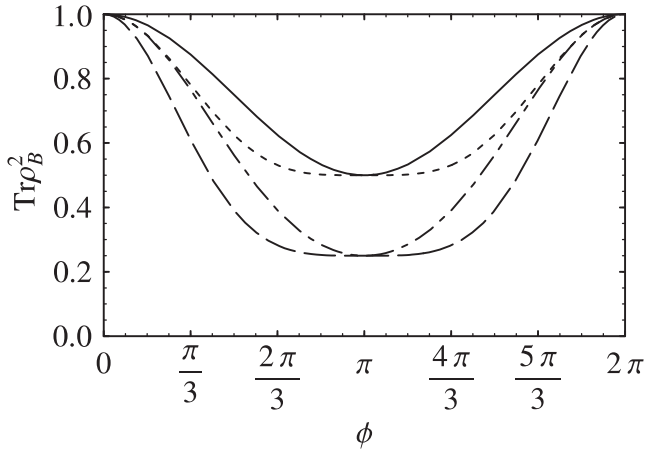}
\caption{Effect of varying $\phi$ on $\pur{\rho}{B}$ where $B$ is
any one atom not at an end (dotted), any two atoms not at ends and
with at least two others between them (dashed), any two or more
consecutive atoms not including an end (dash-dotted), any one or
more consecutive atoms including one end (solid). These purities
are independent of $n$. For $n\geq 6$ corresponding subsets are
e.g.~$B=\{2\},\{2,5\}, \{2,3,4,5\},\{1,2,3,4,5\}$, respectively.
Hence the state is entangled if the purities as listed above are
not in decreasing order that is for all $\phi$ except integer
multiples of $2\pi$.} \label{fig:phics}
\end{figure}

Our network permits the measurement of further correlations and it
does not need the assumption that all atoms have the same single
particle density matrix. It therefore allows a better
characterization of $\ket{\phi_n}$. For any $\phi$ which is not an
integer multiple of $2\pi$, $\ket{\phi_n}$ is a pure state with no
separable subsystems, and hence for any subset $B$ we have
$\pur{\rho}{B}<1=\pur{\rho}{}$ as can be seen from
\figref{fig:phics}. For creating the states $\ket{\phi_n}$ each
qubit only needs to interact with their two nearest neighbors
except for the two extremal atoms 1 and $n$ which will interact
with only their one neighbor. Because of this the reduced purity
of a subset $B$ is determined by the boundary between it and the
rest of the system. Subsets of different sizes but with the same
boundary structure have the same purity. Several examples of these
purities are shown in \figref{fig:phics} as a function of $\phi$.
For example, all sets of two or more adjacent atoms, located
anywhere in the row that do not include either extremal atom have
the same purity $(1+\cos^2(\phi/2))^2/4$ (dash dotted curve in
\figref{fig:phics}) which is independent of $n$. The degree of
violation of \equref{ineq} varies smoothly with $\phi$ and
measuring the various different purities allows to determine
$\phi$ (up to its sign). We will now introduce a quantum network
which detects violations of \equref{ineq} and later, in
Sec.~\ref{secentdectnospat}, show how violations of \equref{ineq2}
can be detected even without achieving spatial resolution of the
different subsystems.

\subsection{Multipartite Entanglement Detection Network}

\begin{figure}[tbp]
\includegraphics[width=8.5cm]{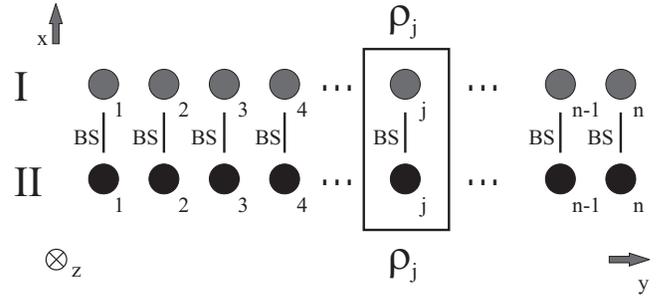}
\caption{Entanglement detection network acting on two rows $I$ and
$II$ of identical qubits. The two rows of $n$ qubits are prepared
in the (possibly entangled) state $\rho_{123\dots n} \otimes
\rho_{123\dots n}$ and performing the pairwise BS followed by
measuring the lattice site occupancies realizes the entanglement
detection network by projecting each operator $\rho_j$ into its
symmetric and antisymmetric subspaces.} \label{BS}
\end{figure}

A family of quantum interferometric networks that directly
estimates $\tr \{\rho^s\}$, $s=2,3,4,\dots$ for any $\rho$ from
$s$ copies of $\rho$ was introduced in \cite{Carolina2002-3}.
These networks rely on the controlled-shift operation $C-V^{(s)}$
between the different copies, where $V^{(s)}\ket{\phi_1}
\ket{\phi_2}\dots \ket{\phi_s}=\ket{\phi_s}\ket{\phi_1}\dots
\ket{\phi_{s-1}}$ for any $\ket{\phi_i}$ with $i=1,2\dots s$. For
the particular case of $s=2$ the value of $\pur{\rho}{}$ is
directly related to the probability of projecting
$\rho\otimes\rho$ into its symmetric and antisymmetric subspaces.
The values of the $2^n$ different purities associated with
$\rho_{123\dots n}$ can hence be determined from the expectation
value of the symmetric and antisymmetric projectors, on each
different pair of reduced states $\rho_{B}=\tr_{A} (\rho_{123\dots
n})$, where $A\cup B=\{1,2,\dots n\}$. These expectation values
can be measured without resorting to the implementation of the
three-qubit C-Swap gate $C-V^{(2)}$ if two identically prepared 1D
rows of $n$ qubits which are represented by bosonic particles are
coupled via pairwise BS, as shown in \figref{BS}
\cite{Carolina2004}.

The BS in the $j$-th column projects the symmetric (antisymmetric)
part of the density operator $\rho_j$ onto doubly (singly)
occupied sites \cite{Bellstateanalyzer} (for details see Appendix
\ref{AppBS}). Two qubits in column $j$ and in state $\rho_{j}
\otimes \rho_{j}$ will thus end up in the same site (+) or in
different sites (-) with probabilities
\begin{equation}
P^{(j)}_{\pm} = \frac{1}{2}\tr\{(\id \pm
V^{(2)})\rho_j\otimes\rho_j\}= \frac{1}{2} \pm
\frac{1}{2}\pur{\rho}{j}.
\end{equation}
Here $S_{\pm}=(\id \pm V^{(2)})/2$ is the symmetric /
antisymmetric projector. By distinguishing doubly occupied sites
from singly occupied ones we can thus determine the purity of
$\rho_{j}$.

Extending this two-qubit scenario to the general case of two
copies of a state of $n$ qubits undergoing pairwise BS (see
Fig.~\ref{BS}) we obtain \cite{Carolina2004}
\begin{equation}
P_{\pm_1\pm_2\dots \pm_n} = \tr\left\{\prod_{i=1}^n\frac{\id \pm_i
V_i^{(2)}}{2}\rho_{12\dots n}\otimes\rho_{12\dots n}\right\}.
\label{prob}
\end{equation}
Inverting the linear equation \equref{prob} the impurity of any
subset of atoms $B$ is given by twice the probability of having an
odd number $j_B$ of antisymmetric projections in subset $B$
\begin{equation}
\pur{\rho}{B} = P(j_B\text{
even})-P(j_B\text{ odd}) = 1- 2P(j_B\text{ odd}). \label{eq:entdet-1s}
\end{equation}
For example, for $n=3$:
\begin{eqnarray}
\pur{\rho}{123} &=& P_{+++} + P_{+--} + P_{-+-} + P_{--+} \\ &&
- P_{---} - P_{+-+} - P_{++-} - P_{-++}, \nonumber \\
\pur{\rho}{12} &=& P_{+++} + P_{++-} + P_{--+} + P_{---} \\ && -
P_{-++} - P_{-+-} - P_{+-+} - P_{+--}, \nonumber \\
\pur{\rho}{1} &=& P_{+++} + P_{++-} + P_{+-+} + P_{+--} \\ && -
P_{-++} - P_{-+-} - P_{--+} - P_{---}. \nonumber
\end{eqnarray}

\subsection{Realization in Optical lattices}
\label{reallatt}

We consider one sheet in the $x-y$ plane of an ultracold
two-species bosonic gas confined in a 3D optical lattice
sufficiently deep so that the system is in a Mott insulating state
with exactly one atom per lattice site \cite{Jaksch98}. Two long
lived internal states of each atom $\ket{a}$ and $\ket{b}$
represent a qubit, i.e. for a site with row index $l$ and column
index $j$ we define two basis states $(a_l^{(j)})^\dagger |{\rm
vac}\rangle \equiv \ket{0}_l^{j}$ and $(b_l^{(j)})^\dagger |{\rm
vac}\rangle \equiv \ket{1}_l^{j}$. Here $|{\rm vac}\rangle$ is the
vacuum state and $a$ ($b$) is the bosonic destruction operator for
an atom in internal state $\ket{a}$ ($\ket{b}$). We assume that
starting from this Mott state identical multipartite entangled
states $\rho_{123\dots n}$ are created in each row while different
rows remain uncorrelated. This can e.g.~be achieved by state
selective cold controlled collisions between atoms in neighboring
columns \cite{BlochEnt, lattice-review}. The above entanglement
detection network can be realized in this setup to study the
entanglement properties of $\rho_{123\dots n}$. We will first
discuss the implementation of the BS by coupling pairs of rows
along the $x$-direction and then investigate methods to
distinguish doubly from singly occupied sites. Note that the
intrinsic parallelism of the 3D optical lattice will allow to run
many copies of these networks in the lattice at once. By
exploiting this parallelism one can obtain an estimate of the
desired projection probabilities and test the violation of
\equsref{ineq}{ineq2} in a single or few experimental runs only.

\subsubsection{The pairwise Beam Splitters}

The dynamics of the atoms in the optical lattice is governed by
the Bose-Hubbard Hamiltonian (BHM) \cite{Jaksch98,
lattice-review}. We assume that any hopping along the $y$ and $z$
directions is suppressed by a sufficient depth of the lattice in
these directions. The hopping in $x$-direction is controlled by
dynamically varying the corresponding lattice depth $V_{0x}$.
Since $V_{0x}$ is proportional to the laser intensity it can
easily be changed in the experiment. The pairwise BS requires that
rows be coupled pairwise (this can be achieved by using a
superlattice of twice the period) and thus we only need to
consider one such pair labelling the two rows by $I$ and $II$,
respectively, as shown in \figref{BS}.

The Hamiltonian describing the dynamics of the atoms in these two
rows can be written as a sum $H_{\rm BHM}= H_{\rm BS} + H_U$ where
$H_{\rm BS}$ is due to hopping along the $x$ direction and $H_U$
is due to the repulsion between two atoms occupying the same
lattice site \cite{Jaksch98, lattice-review}. The two
contributions are given by
\begin{eqnarray}
H_{\rm BS}&=& -J \sum_{j=1}^n (a_I^{(j)\dagger} a_{II}^{(j)} +
b_I^{(j)\dagger} b_{II}^{(j)} + {\rm h.c.} ), \\
 H_{U} &=& \sum_{l=I,II}\sum_{j=1}^N \frac{U}{2}
a_l^{(j)\dagger}a_l^{(j)\dagger} a_{l}^{(j)}a_{l}^{(j)} +
 \nonumber \\ &&
\frac{U}{2} b_l^{(j)\dagger}b_l^{(j)\dagger}b_{l}^{(j)}b_{l}^{(j)}
+ U b_l^{(j)\dagger}b_{l}^{(j)} a_l^{(j)\dagger}a_{l}^{(j)},
\nonumber
\end{eqnarray}
where $J$ is the hopping matrix element and $U$ is the onsite
interaction energy. The parameters $U$ and $J$ depend on the
parameters of the trapping lasers, and their ratio can be varied
over a wide range, on time scales much smaller than the
decoherence time of the system, by dynamically changing the depth
$V_0$ of the optical lattice~\cite{Jaksch98, lattice-review}.

The BS dynamics is perfectly realized in the noninteracting limit
$U=0$ by applying $H_{\rm BS}$ for a time $t_{bs}= \pi/(4J)$ (for
details see Appendix \ref{AppBS}). However, in practice it is
impossible either to control $J$ perfectly accurately or to
completely turn off the interaction $U$, and these imperfections
cause the symmetric component to have a non-zero probability
$q_{bs}$ of failing to bunch which is given by
\begin{equation}
q_{bs} \approx \frac{\pi^2}{8}\left (\frac{\delta
J}{J}\right)^2+\frac{1}{16}\left(\frac{U}{J}\right)^2.
\label{eq:q-bs}
\end{equation}
Here $\delta J$ describe the fluctuations in $J$ (for details see
Appendix \ref{AppBS}). If the fluctuations $\delta J$ occur from
run to run rather than from site to site $q_{bs}$ can be
interpreted as a statistical random variable. We will discuss how
to correct this error in Sec.~\ref{sec:errors}.

\subsubsection{Measuring the lattice site occupation}

Recently, a method that uses atom-atom interactions to distinguish
between singly and doubly occupied sites was demonstrated
experimentally \cite{Measure}. However, a simplified alternative
method where rapid same-site two-atom loss is induced via a
Feshbach resonance and the remaining singly-occupied sites are
detected suffices for our purpose. The detection of the remaining
singly-occupied sites is achieved by measuring the atomic density
profile after the atoms are released from the lattice. A single
Feshbach resonance will cause the loss of either $\ket{aa}$,
$\ket{ab}+\ket{ba}$, or $\ket{bb}$. Hence, in order to empty
doubly-occupied sites in all three states we can either turn on
consecutively three separate Feshbach resonances or change the
internal state of the pairs of atoms during the loss process using
an appropriate sequence of laser pulses. Even if three resonances
are experimentally accessible the latter might yield better
results as the resonance with the best ratio of two-atom to
single-atom loss can be exploited. One suitable sequence
\cite{Foot1} which does not require precise control is to apply a
large number of pulses of random relative phase and approximate
area $\pi/2$ each at equal intervals. Each initial state then
spends $1/3$ of the time in the resonant state, and hence has
probability $q_l= \exp(-t_l/3\tau_d)$ of failing to lose a pair of
atoms occupying the same site where $t_l$ is the total duration of
the sequence and $\tau_d$ the two-atom loss time constant. This
method will not be perfect since single particles are also lost
from the system with some time constant $\tau_s$ and hence $t_l$
cannot be chosen arbitrarily large. The probability of losing a
single particle is $p_l = 1-\exp(-t_l/\tau_s)$ where $\tau_s \gg
\tau_d$. Both $p_l$ and $q_l$ are error probabilities, and their
sum $p_l+q_l$ is minimized by choosing $t_l=\ln(\tau_s/(3\tau_d))/
(1/3\tau_d-1/\tau_s)$. An experiment recently performed by Widera
\textit{et. al} \cite{Measure} measured $\tau_d=1.3{\rm ms}$ and
no detectable loss of single atoms for resonance times up to
$100{\rm ms}$. If we take $\tau_s=500{\rm ms}$ the above gives
$t_l=19 \rm ms$ and error probabilities of $q_l=0.8\%$ and
$p_l=3.7\%$.

In summary, the optical lattice realization of our entanglement
detection network contains the following stages at which
experimental errors are likely to occur: (i) $q_{bs}$ from the
implementation of $H_{BS}$; (ii) $p_l$ and $q_l$ occurring during
the loss stage; (iii) detector errors $p_d$ in counting the number
of singly occupied lattice sites. In addition (iv) the setup might
lack spatial resolution in atom counting. We consider how to
correct errors (i) - (iii) in Sec.~\ref{sec:errors} and next study
the case (iv) of no spatial resolution in an otherwise perfect
experimental setup.

\section{Entanglement detection and characterization without
spatial resolution} \label{secentdectnospat}

We assume that the measurement of the total number of
singly/doubly occupied sites is accurate but that we cannot know
their locations. We first show that this information is sufficient
to detect a violation of \equref{ineq2}. Then we study how various
different experimentally realizable multipartite entangled states
might be characterized using such measurements.

\subsection{Entanglement detection}

The probabilities $P(j)$ of measuring $j$ singly occupied sites in
one row ($2j$ single atoms in total) are given by
\begin{eqnarray}
P(j)&=&\sum\limits_{|A|=j}\frac{1}{2^n}\sum\limits_B (-1)^{|A\cap
B|}\pur{\rho}{B}\\&=&\frac{1}{2^n}\sum\limits_{k=0}^n{n\choose
k}\avpur{\rho}{k}\sum\limits_l{k\choose l}{n-k\choose
j-l}(-1)^l, \nonumber
\end{eqnarray}
where the summation indices are $k=|B|$, $l=|A\cap B|$, $j=|A|$,
and $A$ is the set of singly occupied (antisymmetric) sites. We
form the generating function
\begin{eqnarray}
\sum\limits_{j=0}^nx^jP(j)&=&\frac{1}{2^n}\sum\limits_{k=0}^n{n\choose
k}\avpur{\rho}{k}(1-x)^k(1+x)^{n-k}\nonumber\\&=&\left(\frac{1+x}{2}\right)^n\sum\limits_{k=0}^n{n\choose
k}\avpur{\rho}{k}\left(\frac{1-x}{1+x}\right)^k \nonumber \\
\end{eqnarray}
and let $y=(1-x)/(1+x)$, to obtain
\begin{equation}
\sum\limits_{j=0}^n(1-y)^j(1+y)^{n-j}P(j)=\sum\limits_{k=0}^ny^k{n\choose
k}\avpur{\rho}{k},
\end{equation}
from which we find
\begin{equation}
\avpur{\rho}{k}=\left[{n\choose
k}\right]^{-1}\sum\limits_{j=0}^nP(j)\sum\limits_l{j\choose
l}{n-j\choose k-l}(-1)^l \label{eq:entdet-1}.
\end{equation}
Therefore, although we cannot determine the purity of a given
subset of the row of atoms we can still determine average purities
associated with subsets of atoms of a given size by measuring
$P(j)$. We will prove later (see \equref{Vbound-q}) that the
accuracy in finding $P(j)$ required for obtaining a given accuracy
of $\avpur{\rho}{k}$ has an upper bound independent of $n$ and $k$
if no errors are present. The network is thus efficient in
detecting the presence of entanglement in all pure (and some
mixed) entangled states via the violation of \equref{ineq2}.

\begin{figure}
\includegraphics[width=8.5cm]{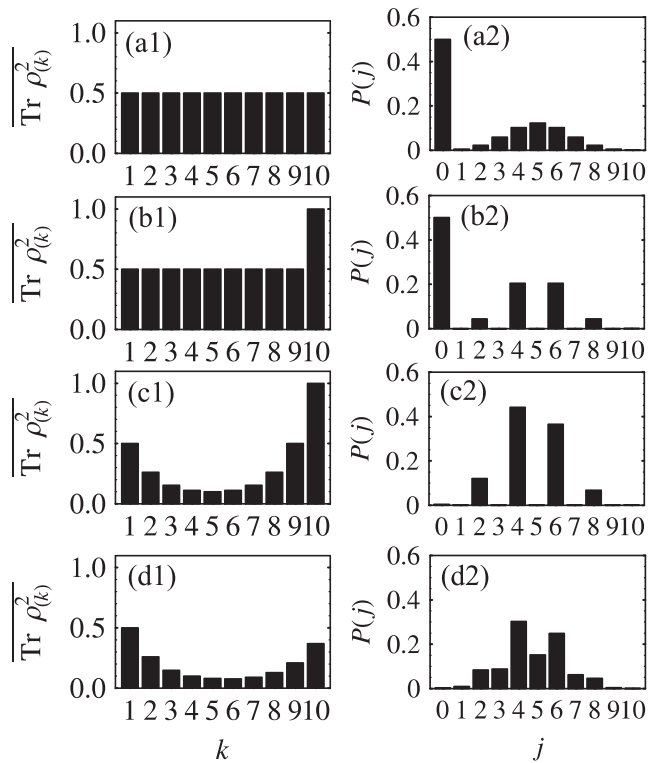}
\caption{$\avpur{\rho}{k}$ against $k$ (left) and $P(j)$ against
$j$ (right) for $n=10$. (a) Classically correlated state
$\rho=(\ket{0}^n\bra{0}^n+\ket{1}^n\bra{1}^n)/2$. (b) Maximally
entangled state $\ket{(\gamma=0)_{10}}$. (c) Cluster state
$\ket{(\phi=\pi)_{10}}$. (d) Same as (c) with 10\% dephasing
decoherence.} \label{fig:kjplots}
\end{figure}

In \figref{fig:kjplots} we show the probabilities $P(j)$ and the
resulting average purities for a variety of different states. For
a classically correlated state shown in \figref{fig:kjplots}a the
values of $\avpur{\rho}{k}$ are monotonically decreasing with $k$
showing that \equref{ineq2} is not violated. The maximally
entangled state shown in \figref{fig:kjplots}b has the
characteristic that $\avpur{\rho}{k}=1/2$ for $0<k<n$ while
$\avpur{\rho}{n}=1$ and thus the inequalities are violated in this
case. The cluster state shown in \figref{fig:kjplots}c violates
the inequalities for all $j>n/2$ and therefore its entanglement is
detected. Finally, in \figref{fig:kjplots}d we show a noisy
cluster state which was affected by phase noise acting
independently on each atom. It can clearly be seen that
decoherence reduces the violation of the inequalities but that
entanglement is detectable for small amounts of noise.

\subsection{Characterization of entanglement}

The $n$ measurable quantities $\avpur{\rho}{k}$, $k=1,\dots ,n$ do
not provide us with enough information to determine an arbitrary
state $\rho$. However, if it may be assumed that the state $\rho$
is of a known form with less than $n$ unknown parameters, it is
often possible to determine these parameters from
$\avpur{\rho}{k}$. We demonstrate this by considering macroscopic
superposition states $\ket{\gamma_n}$ and cluster-like states
$\ket{\phi_j}$ introduced in Sec.~\ref{subsec:CharId}. Finally, we
will also look at product states of states of subsystems
containing several atoms.

\subsubsection{Macroscopic superposition states}

Because the state $\ket{\gamma_n}$ is totally symmetric the
individual purities given in \equref{macropur} only depend on the
size of the subsystem $k=|B|$ and thus $\avpur{\rho}{k} =
\pur{\rho}{B}$. Hence, from the knowledge of $\avpur{\rho}{k}$ we
can determine the value of $\gamma$ which in principle only
requires two of the purities. The remaining equations allow a
partial check of the assumption that the measured state indeed has
the form $\ket{\gamma_n}$.

\subsubsection{Cluster states}

\begin{figure}
\includegraphics[width=8.5cm]{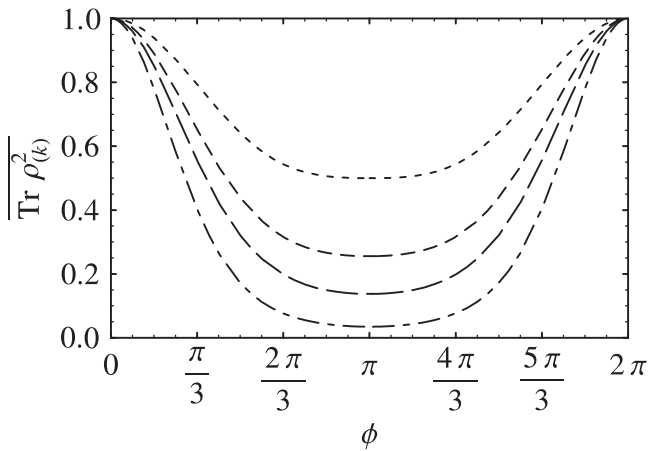}
\caption{Average purities $\avpur{\rho}{k}$ ($k=1$ dotted curve,
$k=2$ dashed curve, $k=3$ dash-dotted curve, $k=7$ solid curve)
for the state $\ket{\phi_{n=15}}$. Since all states are pure we
have $\avpur{\rho}{k}= \avpur{\rho}{n-k}$.} \label{fig:phi15c}
\end{figure}

The states $\ket{\phi_n}$ are parameterized by the entangling
phase $\phi$. The average purities as a function of $\phi$ are
depicted in Fig.~\ref{fig:phi15c}. For any value of $0 <\phi< 2
\pi$ the states violate the inequalities \equref{ineq2}. The
degree of violation increases with $\phi$ until $\phi=\pi$ where
the state is a cluster state and the degree of violation is a
maximum. If $\phi$ is increased further the state again approaches
a product state and the degree of violation of the inequalities
correspondingly decreases. Hence, from experimentally measured
$\avpur{\rho}{k}$ one can determine the phase $\phi$ up to its
sign. Again the over determined system ($n$ equations for one
unknown $\phi$) provides a check on how well the state fits the
assumed form $\ket{\phi_n}$.

\begin{figure}
\includegraphics[width=8.5cm]{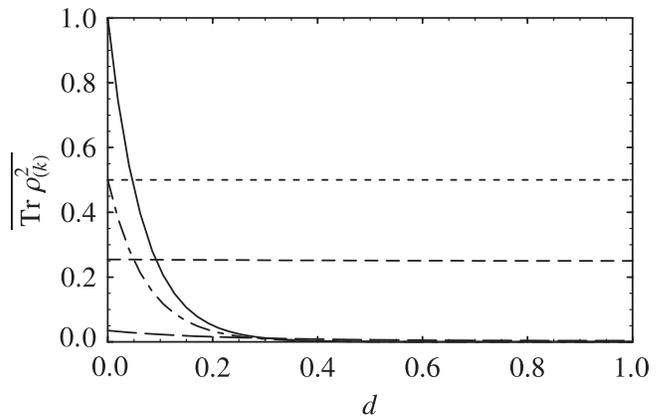}
\caption{Effect of dephasing noise on the $n=15$ cluster state,
$k=1$ (dotted), $k=2$ (short dashed), $k=8$ (long dashed), $k=14$
(dash-dotted), $k=15$ (solid).} \label{fig:cluster15deph}
\end{figure}

The effect of dephasing according to the map \equref{phasenoise}
on a cluster state is shown in \figref{fig:cluster15deph}. The
average purities decrease with increasing noise level $d$.
Entanglement is certainly present and in principle detectable by
our method as long as the $\avpur{\rho}{k}$ are not in descending
order, in the case shown in \figref{fig:cluster15deph} up to
$d\approx 0.45$. Again, the parameter $d$ can in principle be
determined from measuring the average purities.

\subsubsection{Products of entangled subsystem states}

Finally we give an example of a class of states where even though
$\rho$ is characterized by more than $n$ parameters, the
associated average purities $\avpur{\rho}{k}$ only depend on $n$
parameters. Consider the case where $\rho$ is a product state of
$L$ subsystems, $\rho=\otimes_{i=1}^L\rho_i$, with each subsystem
$\rho_i$ composed of a known number $n_i$ of atoms. In this case
we have
\begin{equation}
\avpur{\rho}{k} = \sum
\limits_{\stackrel{\scriptstyle\{k_1,\ldots,k_j\}}
{\scriptstyle\sum_i k_i=k}}\prod_i{n_i\choose
k_i}\overline{\pur{\rho}{i\left(k_i\right)}}.
\end{equation}
Since there are $\sum_i n_i=n$ different
$\overline{\pur{\rho}{i\left(k_i\right)}}$ in total,
$\avpur{\rho}{k}$ provides us with enough information to determine
the average purities $\overline{\pur{\rho}{i\left(k_i\right)}}$ of
every subsystem. In particular, $L=n/2$ and all $n_i=2$ is the
case of only pairwise entanglement. Since \equref{purbounds} holds
for all $\rho_i$ taking $n_i=2$ provides a test for multi-particle
as opposed to two-particle correlations in a given state. We note
that unless $\rho$ is pure classical multipartite correlations
will also be detected by our network.

We will now study how the working of the network is affected by
experimental errors. In particular we will estimate how many runs
are necessary to obtain the probabilities $P(j)$ with sufficient
accuracy in the presence of errors.

\section{Effects of experimental error}
\label{sec:errors}

The errors introduced in Sec.~\ref{reallatt} affect the ability to
find the purities $\pur{\rho}{B}$ as well as the average purities
$\avpur{\rho}{k}$ associated with $\rho_{12\dots n}$. All of these
errors are of one of two mathematical kinds: extra \emph{pairs} of
atoms and missing \emph{single} atoms. We call these two errors
``beam-splitter'' and ``detector'' error respectively. Their
respective probabilities $q=q_{bs}+q_l$ and $p=p_d+p_l$ are
understood to include also errors occurring while particles are
lost from doubly occupied sites. The relationship
\equref{eq:entdet-1s} between the purities of $\rho_{B}$ and the
probabilities of an even/odd number $j_B$ of singly occupied sites
in $B$ indicates that an experimental error, occurring with
probability $p$ per site, changes the result $\pur{\rho}{B}$ by
$O(|B|p)$ if we do not attempt to correct for it. This renders the
measured results totally meaningless as soon as $|B|p\sim 1$,
because the purity of any given state $\rho_B$ is smaller or equal
to one. However certain types of error, including the BS and
detector errors, can be corrected by a suitable modification of
the formulas \equsref{eq:entdet-1s}{eq:entdet-1} yielding
$\pur{\rho}{B}$ and $\avpur{\rho}{k}$. This correction eliminates
systematic errors, making $\pur{\rho}{B}$ and $\avpur{\rho}{k}$
correct on average, but tends to amplify the random errors that
are inevitable in measuring probabilities using a finite number of
experimental runs. These random errors can in principle be made
arbitrarily small for any $p,q<1$ by increasing the number of
runs, but in practice there is a limit because, as we will show,
the number of runs required scales approximately exponentially in
$|B|p$. We will now investigate the effects of these errors on the
performance of the entanglement detection network both without and
with spatial resolution.

\subsection{Without spatial resolution}

We assume the probabilities $p$ and $q$ to be the same for all
$2n$ lattice sites and in the case of $q$ for all symmetric atom
pair states $\ket{aa}$, $\ket{bb}$ and $\ket{ab}+\ket{ba}$. Errors
at different lattice sites are assumed to be uncorrelated.

\subsubsection{Beam splitter error}
\label{BSE}

Let $P_{\rm exp}(i)$ be the probability of detecting $i$ atoms in
an experimental run. If only BS errors are present this
probability is given by
\begin{equation}
P_{\rm exp}(2i)=\sum\limits_{j=0}^iP(j){n-j\choose
n-i}q^{i-j}(1-q)^{n-i}, \label{eq2i}
\end{equation}
where the factor two in $P_{\rm exp}(2i)$ accounts for $P(j)$
being the probability of having $j$ antisymmetric pairs. We can
use generating functions to invert \equref{eq2i}
\begin{equation}
\sum\limits_{i=0}^{n}x^iP_{\rm
exp}(2i)=\sum\limits_{j=0}^nP(j)x^j(1-q+xq)^{n-j},
\end{equation}
leading to
\begin{equation}
P(j)=\sum\limits_{i=0}^j{n-i\choose
n-j}\frac{(-q)^{j-i}}{(1-q)^{n-i}}P_{\rm exp}(2i)
\label{eq:entdet-corrector2}.
\end{equation}
We now apply \equref{eq:entdet-corrector2} to a subsystem $B$ and
substitute this into \equref{eq:entdet-1s}, giving
\begin{equation}
\pur{\rho}{B}=\sum\limits_{i_B=0}^{k}\left(\frac{1+q}{1-q}\right)^{k-i_B} (-1)^{i_B} P_{\rm exp}(2i_B),
\label{eq:entdet-1qs}
\end{equation}
where $i_B$ refers to the number of atoms detected in $B$. This
expression is then averaged over all $B$ of size $|B|=k$ to give
\begin{equation}
\avpur{\rho}{k}=\sum\limits_{i=0}^{n}A_{ki} P_{\rm exp}(2i),
\end{equation}
with
\begin{equation}
A_{ki}=\left[{n\choose
k}\right]^{-1}\sum\limits_{l=0}^k(-1)^l\left(\frac{1+q}{1-q}\right)^{k-l}{i\choose
l}{n-i\choose k-l}. \label{eq:entdet-corrector2a}
\end{equation}

Hence, using \equref{eq:entdet-corrector2a} instead of
\equref{eq:entdet-1} corrects all the \emph{systematic} error
caused by an imperfect BS. We are still left with the inherent
\emph{random} error associated with the measurement of the
probabilities $P_{\rm exp}(2i)$, which is reduced by increasing
the number of experimental runs.

Because of this random error the estimate of $\avpur{\rho}{k}$
obtained from $N$ experimental runs (each using one pair
$\rho\otimes\rho$) has the correct mean but a nonzero standard
deviation $\sqrt{V_k/N}$, where this defines $V_k(p,q,\rho)$;
hence $O(V_k)$ runs are necessary for meaningful results. Note
that in general $V_k>0$ even when $p=q=0$, as it includes the
inherent quantum uncertainty as well as that added by experimental
error. In the case of BS error
\begin{eqnarray}
V_k&=&\sum_iP_{\rm exp}(i)A_{ki}^2- \left(\avpur{\rho}{k}\right)^2 <
{\rm max}_i \left(A_{ki}^2\right) \nonumber \\
&\leq& \left(\frac{1+q}{1-q}\right)^{2k} \approx
e^{4kq}\label{Vbound-q},
\end{eqnarray}
where the approximation is valid for $k \gg 1$, $q \ll 1$. The
bound \equref{Vbound-q} proves that the number of runs required to
obtain meaningful estimates of $\avpur{\rho}{k}$ is reasonable for
$k\lesssim 1/q$, however large $n$ is.

\begin{figure}
\includegraphics[width=8.5cm]{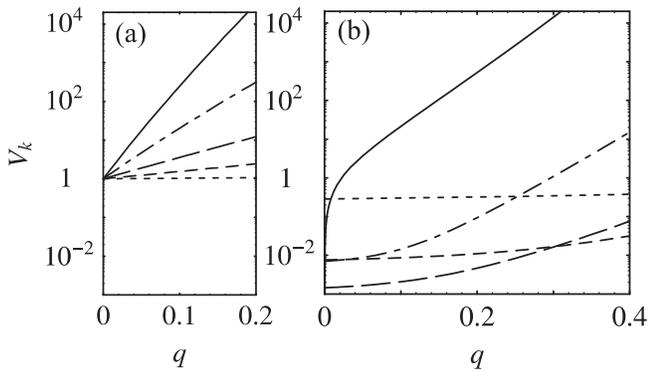}
\caption{Variance $V_k$ against BS error, for $k=1$ (dotted),
$k=4$ (short dashed), $k=7$ (long dashed), $k=11$ (dash-dotted),
$k=15$ (solid) curve, and $n=15$. (a) shows the result for the
worst case and (b) for a cluster state.} \label{fig:array15q}
\end{figure}

\begin{figure}
\includegraphics[width=8.5cm]{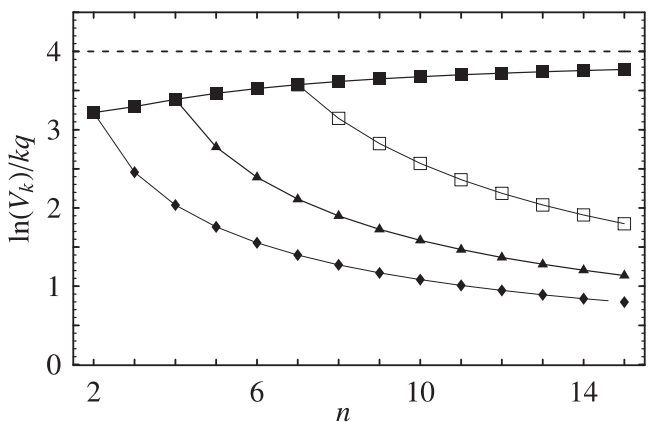}
\caption{Worst case variance against $n$ with BS error $q=1/k$,
for $k=2$ (diamonds), $k=4$ (triangles), $k=7$ (hollow squares),
$k=n$ (solid squares). The dotted horizontal line is the analytic
bound $V_k\leq \exp(4kq)$. The solid curves are drawn to guide the
eye.} \label{fig:max15qn}
\end{figure}

We numerically computed the worst case by maximizing $V_k$ with
respect to $\avpur{\rho}{k}$ subject only to \equref{purbounds}
and compare it to cluster states $\ket{(\phi=\pi)_n}$ in
\figref{fig:array15q}. The results confirm the analytically found
exponential increase of $V_k$ with $q$ in the worst case. For the
cluster state $V_k$ increases only slowly with $q$ for small $k
\leq 1/q$ while for $k \gtrsim 1/q$ we find approximately
exponential growth of $V_k$ with $q$. We also computed the
variances for maximally entangled states and found that they are
quite close to the worst case shown in \figref{fig:max15qn}a.
Therefore one may in an experiment generally not expect the
variances $V_k$ to be much smaller than our worst case results.
Thus only BS errors up to $q=1/k$ are acceptable and yield
reliable results in a reasonable number of runs for all $0\leq k
\leq n$. However, as shown in \figref{fig:max15qn} for $q=1/k$,
the average purities with $k \ll n$ will be determined much more
accurately than those with $k \approx n$ and should thus be
preferentially used for determining parameters characterizing the
measured state.

We finally note that if $J$ is fluctuating from run to run and
$q_{bs}$ hence becomes a random variable error correction is still
possible. In this case we have to replace \equref{eq2i} by
\begin{eqnarray}
P_{\rm exp}(2i)&=&\sum\limits_{j=0}^iP(j){n-j\choose n-i} \times
\nonumber \\
&&\quad \int {\rm d} J f(J) q(J)^{i-j}(1-q(J))^{n-i},
\label{ranvarj}
\end{eqnarray}
where $f(J)$ is the probability density function of $J$ and $q(J)$
denotes the BS error as a function of $J$. The resulting system of
linear equations \equref{ranvarj} can be treated using the methods
introduced above.

\subsubsection{Detector error}

\begin{figure}
\includegraphics[width=8.5cm]{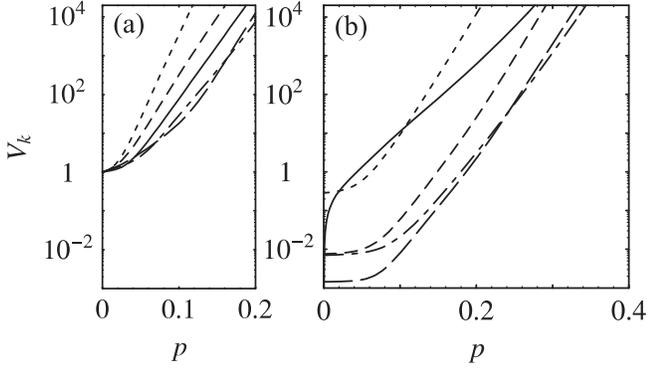}
\caption{Variance against detection error without spatial
resolution using \equref{eq:entdet-corrector1}, for $k=1$
(dotted), $k=4$ (short dashed), $k=7$ (long dashed), $k=11$
(dash-dotted), $k=15$ (solid), and $n=15$. (a) shows the result
for the worst case and (b) for a cluster
state.}\label{fig:array15p}
\end{figure}

\begin{figure}
\includegraphics[width=8.5cm]{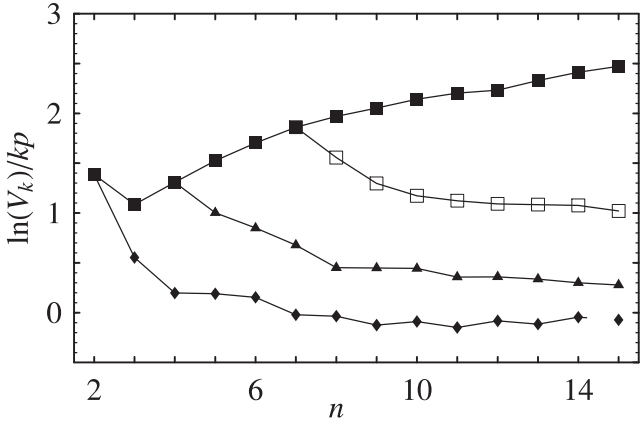}
\caption{Worst case variance against $n$ with detection error
$p=1/(2k)$ using least squares, for $k=1$ (diamonds), $k=2$
(triangles), $k=3$ (hollow squares), $n$ (solid squares). The
solid curves are drawn to guide the eye.} \label{fig:max15lsn}
\end{figure}

We now assume that only detector errors are present and each atom
has a probability $p$ of failing to be detected. In this case
$P_{\rm exp}(i)$ is related to $P(j)$ via
\begin{equation}
P_{\rm exp}(i)=\sum\limits_{j=i/2}^nP(j){2j\choose
i}p^{2j-i}(1-p)^i. \label{eq:errorp}
\end{equation}
We can again solve this equation by methods similar to those used
in Sec.~\ref{BSE} (though this time it is not a true inverse
because the system is over-determined \cite{Foot2}) to obtain
\begin{equation}
P(j)=\sum\limits_{i=2j}^n{i\choose
2j}\frac{(-p)^{i-2j}}{(1-p)^i}P_{\rm exp}(i),
\label{eq:entdet-corrector1}
\end{equation}
where non-integer values of $j$ are discarded. By combining
\equsref{eq:entdet-1}{eq:entdet-corrector1} we obtain
$\avpur{\rho}{k}$ in terms of $P_{\rm exp}(i)$. For the remaining
random error measured by $V_k$ we obtain the upper bound
\begin{equation}
V_k\leq \left(\frac{1+p}{1-p}\right)^{4n}\approx e^{8np}.
\label{eq:anabound}
\end{equation}
Numerical calculations for $n\leq 15$ confirm this exponential
growth of $V_k$ with $p$ at $np\sim 1$. The results are shown in
\figref{fig:array15p}. This time, however, $V_k$ is typically much
smaller than the analytic bound, e.g.~for the $n=15$ cluster
state, fitting $V_k\propto\exp(\beta np)$ gives $\beta\approx 2$.
However, the exponential growth with $n$ implies a practical limit
of $n\sim 1/p$ for any $k$ contrary to the case of BS errors. This
scaling can be improved by using the least squares method to
handle the over-determined linear system of equations
\equref{eq:errorp}. We do not have an analytic bound analogous to
\equref{eq:anabound} for the least squares method but numerical
calculations for the cluster state and the worst case result in
significantly lower values for $V_k$ than those obtained from
\equref{eq:entdet-corrector1}. Most importantly as shown in
\figref{fig:max15lsn} it appears that the scaling of $V_k$ becomes
exponential in $kp$ rather than $np$ similarly to the case of BS
errors. This implies that an error of $p \lesssim 1/k$ is
acceptable for obtaining meaningful estimates of $\avpur{\rho}{k}$
in a reasonable number of experimental runs.

In an actual experiment both BS and detector errors will be
present and first correcting for the detector error $p$ using
\equref{eq:entdet-corrector1}, then substituting the resulting
$P(j)$ for $P_{\rm exp}(2j)$ in \equref{eq:entdet-corrector2a} to
correct for the BS error $q$ yields a combined analytical error
bound of
\begin{equation}
V_k\leq
\left(\frac{1+p}{1-p}\right)^{4n}\left(\frac{1+q}{1-q}\right)^{2k}
\approx e^{8np+4kq}.
\end{equation}
According to our numerical results using the least squares method
this bound can be improved requiring only $p,q\lesssim 1/k$ for
obtaining sufficiently small errors.

\subsection{With spatial resolution}

If spatial resolution is available then $\pur{\rho}{B}$, not just
its average $\avpur{\rho}{k}$ over all subsets $B$ of size $k$,
becomes accessible to measurement. This allows us to do some state
characterizations which would otherwise be impossible and also
introduces extra redundancy. As we will show below this does not
affect the tolerance to BS errors but we will find that the
detector error tolerance improves to $p^2\sim 1/|B|$.
Imperfections in the spatial resolution, however, will lead to
additional errors in determining $\pur{\rho}{B}$.

\subsubsection{Beam splitter error}

The variance of $\pur{\rho}{B}$ due to BS errors can be directly
inferred from \equref{eq:entdet-1qs}, where only the atoms in $B$
are counted. Using the same methods as in Sec.~\ref{BSE} we find
an upper bound for $V_B$ (defined analogously to $V_k$) given by
$V_B \leq ((1+q)/(1-q))^{2|B|}\approx \exp(4|B|q)$. The BS error
tolerance (for fixed $k=|B|$) is thus independent of whether
spatial resolution is available or not.

\subsubsection{Detector error}

The situation is different for the detector error. Each
antisymmetric pair contains two atoms only one of which needs to
be detected to know that it was antisymmetric. Therefore the
effective error probability becomes $p^2$. The resulting formula
is
\begin{equation}
\pur{\rho}{B}=\sum\limits_{i=0}^{|B|}(-1)^i
\left(\frac{1+p^2}{1-p^2}\right)^{i}P_{\rm exp}^B(i)
\label{eq:entdet-q2s},
\end{equation}
where $P_{\rm exp}^B(i)$ is the probability of measuring $i$
antisymmetric sites in subset $B$. The variance bound is given by
$V_B \leq ((1+p^2)/(1-p^2))^{2|B|} \approx \exp(4|B|p^2)$.

\subsubsection{Imperfect spatial resolution}

There is a new type of error to consider as the spatial resolution
itself will not in practice be perfect. If we let $f(x,y)$ be the
probability of finding at position $x$ a particle which is
actually at position $y$, then
\begin{equation}
P_{\rm exp}(A)=\sum\limits_{i \in B}
\sum\limits_\varsigma\frac{1}{s(A)}
\prod\limits_{i=0}^{k}f(A_{\varsigma(i+k)},B_i)
f(A_{\varsigma(i)},B_i)P(B)\label{eq:res-error},
\end{equation}
where $A$ is the ``set'' (its elements are not necessarily
distinct) of observed atom positions, and $P_{\rm exp}(A)$ is the
experimental probability of observing atoms exactly at these
positions $A$. The set $B$ denotes the antisymmetric sites and the
probability of having antisymmetric sites at positions $B$ is
$P(B)$, $k=|B|=|A|/2$, and $\varsigma$ runs over all $(2k)!$
permutations of the $2k$ atoms in $A$. The $i$-th element of $B$
and $A$ are written as $B_i$ and $A_i$, respectively. The symmetry
factor $s(A)$ stands for the number of permutations $\varsigma$
which leave the ordered lists of atoms $\{A_i\}$ invariant (e.g.,
$s(\{1,1\})=2$, $s(\{1,2\})=1$), and is needed because our
summation runs over different ordered lists $\{A_i\}$ of the same
set $A$. This is an over-determined linear system, and just as in
the case of detector error, we can either explicitly solve it by
discarding some of the equations or apply the least squares
method. As an explicit solution we can e.g.~use
\begin{equation}
P(B) = \sum\limits_{\{A_i\}}\frac{s(A)}{2^k}
\left(\prod\limits_{i=1}^{2k}f^{-1}(B_i,A_i)\right)P_{\rm
exp}(A)\label{res-corrector},
\end{equation}
where we define $B_{k+i}\equiv B_i$. The sum runs over all ordered
lists $\{A_i\}$ and $f^{-1}$ is the matrix inverse of $f$, that is
$f^{-1}(y,x)f(x,z)= \delta_{xz}$.

\begin{figure}
\includegraphics[width=8.5cm]{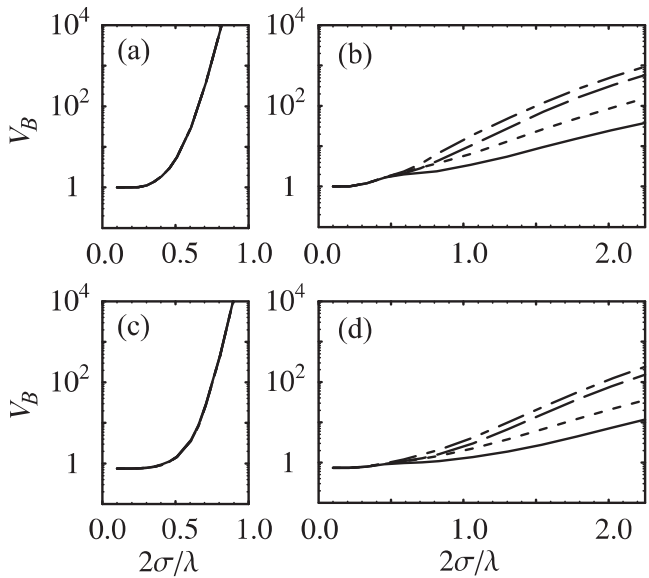}
\caption{Variance of $\pur{\rho}{B}$ as a function of the standard
deviation $\sigma$, for $B=\{2\}$ (dotted), $B=\{2,3\}$ (dashed),
$B=\{1,3\}$ (dash-dotted), $B=\{1,2,3\}$ (solid), and $n=4$. (a)
and (b) are worst case variances while (c) and (d) are for a
maximally entangled state. (a) and (c) each show four almost
coincident curves obtained from \equref{res-corrector} while (b)
and (d) are found by the least squares method.}
\label{fig:array4s}
\end{figure}

As before by performing numerical calculations using the least
squares method we find much lower variances than with
\equref{res-corrector}. An example is shown in
\figref{fig:array4s} where we plot the variance of $\pur{\rho}{B}$
due to a Gaussian position error of the form
\begin{equation}
f(x,y)=\frac{1}{\sqrt{2\pi}\sigma}
\int\limits_{x-\lambda/4}^{x+\lambda/4} {\rm d}z
\exp[-(z-y)^2/(2\sigma^2)] \label{poserr}
\end{equation}
where $\sigma$ is the standard deviation and $\lambda$ is the wave
length of the laser creating the optical lattice. The
corresponding lattice spacing is $\lambda/2$. The results obtained
from \equref{res-corrector} shown in \figref{fig:array4s}a, c
require a resolution of $\sigma \lesssim \lambda/2$ whereas the
least squares method shown in \figref{fig:array4s}b, d yields
reasonable variances $V_B$ for spatial resolutions up to $\sigma
\lesssim 3 \lambda /2$. However, due to the exponentially large
number of possibilities for $A$ the least squares calculation
becomes intractable for large $n$.

\section{Conclusion}
\label{secconcl}

We discussed the detection and characterization of multipartite
entanglement in optical lattices with the quantum network
introduced in \cite{Carolina2004} under different experimental
conditions. We first described how the network can be implemented
in ideal experimental conditions and showed that it allows to
characterize a number of important classes of states like
cluster-like states and macroscopic superposition states. We
investigated the experimental realization of the network in an
optical lattice and identified lack of spatial resolution, errors
in the BS operation and imperfect atom detection as the main
sources of error. We showed that even in the absence of spatial
resolution entanglement can be detected. In cases where the
entangled state is characterized by a few parameters only we found
that these can often be determined from the measurement results.
We also studied the influence of BS errors occurring with
probability $q$ and detection errors with probability $p$ and
concluded that with small numbers of experimental runs
entanglement can be detected between $k$ atoms as long as $k p
\lesssim 1$ and $k q \lesssim 1$. Finally we showed that for
obtaining purities $\pur{\rho}{B}$ of subsets $B$ rather than
average purities with reasonable experimental effort a spatial
resolution of $\sigma \approx \lambda$ is necessary.

The results obtained in this work show that unambiguous
multipartite entanglement detection in optical lattices is
possible with current technology. This has not yet been achieved
experimentally \cite{BlochEnt}. Furthermore it will even be
possible to determine some of the characteristics of entangled
states created in these experiments without the requirement of
performing spatially resolved measurements. Our network is thus a
viable alternative to detecting entanglement via witnesses or full
quantum state tomography.

\begin{acknowledgments}
This work was supported by EPSRC through the QIP IRC
(www.qipirc.org) GR/S82176/01 and project EP/C51933/1, and by the
EU network OLAQUI. C.M.A. thanks Artur Ekert for useful
discussions and is supported by the Funda{\c c}{\~a}o para a
Ci{\^e}ncia e Tecnologia (Portugal).
\end{acknowledgments}

\appendix

\section{The beam splitter operation}
\label{AppBS}

Since the BS only couples two lattice sites in rows I and II of
each column (see \figref{BS}) we consider a single such pair and
omit the column superscript $j$ in this section. For $U=0$, we
obtain from the Heisenberg equations for the operators $\alpha
=a,b$
\begin{eqnarray}
 \alpha_I(t) &=& \cos(J t) \alpha_I -i \sin(J t)\alpha_{II},  \nonumber \\
 \alpha_{II}(t) &=& \cos(J t) \alpha_{II} -i \sin(J t)\alpha_I.
\end{eqnarray}
Hence, applying $H_{\rm BS}$ for a time $t_{bs}= \pi/(4J)$
implements a perfect pairwise BS.

Initially the atom pair is in a state of the form $\rho \otimes
\rho$, where $\rho$ is a single qubit state and hence has a
spectral decomposition of the form $\rho=\lambda_1 \ket{c}\bra{c}
+ \lambda_2 \ket{d}\bra{d}$, with $\lambda_1 + \lambda_2=1$,
$\ket{c}=c^\dagger \ket{\rm vac}$, and $\ket{d}=d^\dagger \ket{\rm
vac}$. Here $c^\dagger$, $d^\dagger$ are linear superpositions of
$a^\dagger$, $b^\dagger$ with coefficients depending on $\rho$.
Therefore we can write
\begin{eqnarray}
\rho \otimes\rho&=& \lambda_1^2 \ket{c_Ic_{II}}\bra{c_Ic_{II}} +
\lambda_2^2 \ket{d_Id_{II}}\bra{d_Id_{II}} \\ && +
\frac{\lambda_1\lambda_2}{2}
(\ket{c_Id_{II}}+\ket{d_Ic_{II}})(\bra{c_Id_{II}}+\bra{d_Ic_{II}})\nonumber\\&&
+\frac{\lambda_1\lambda_2}{2}
(\ket{c_Id_{II}}-\ket{d_Ic_{II}})(\bra{c_Id_{II}}-\bra{d_Ic_{II}}),
\nonumber
\end{eqnarray}
which is a classical mixture of a symmetric state with probability
$P_+ = 1-\lambda_1\lambda_2$ and an antisymmetric state with
probability $P_- = \lambda_1\lambda_2$. After the BS the resulting
state $\rho'=\exp(iH_{BS}t)\rho \otimes \rho \exp(-iH_{BS}t)$ is
given by
\begin{eqnarray}
\rho' &=& \lambda_1^2 \ket{\Phi_1}\bra{\Phi_1} + \lambda_2^2
\ket{\Phi_2}\bra{\Phi_2} \\ && + \lambda_1\lambda_2
\ket{\Phi_3}\bra{\Phi_3} + \lambda_1\lambda_2
\ket{\Phi_4}\bra{\Phi_4}, \nonumber
\end{eqnarray}
where $\ket{\Phi_1} = (c_I^\dagger c_{I}^\dagger + c_{II}^\dagger
c_{II}^\dagger) \ket{\rm vac}/2$, $\ket{\Phi_2} = (d_I^\dagger
d_{I}^\dagger + d_{II}^\dagger d_{II}^\dagger) \ket{\rm vac}/2$,
$\ket{\Phi_3} = (c_I^\dagger d_{I}^\dagger + c_{II}^\dagger
d_{II}^\dagger) \ket{\rm vac}/\sqrt{2}$ are states with double
occupancy in one row and an empty site in the other row while
$\ket{\Phi_4} = (c_I^\dagger d_{II}^\dagger - c_{II}^\dagger
d_{I}^\dagger) \ket{\rm vac}/\sqrt{2}$ is a state with a singly
occupied site in each row. Hence, after the BS we will find a
doubly occupied site with probability $1-\lambda_1\lambda_2=P_+$
while two singly occupied sites result with probability
$\lambda_1\lambda_2=P_-$.

However, in practice it is impossible to completely turn off the
interaction $U$ which may result in a symmetric state failing to
bunch. We consider any symmetric state $\ket{\Psi_s}$ and because
$H_{BS}+H_U$ acts only on the row indices, not the internal
states, and is symmetric between the two rows the probability
$q_{bs}$ of failure to bunch is given by
\begin{eqnarray}
q_{bs}&=& |\bra{\Psi_s}e^{i(H_{BS}+H_U)t}\ket{\Psi_s}|^2 \\
&=&\frac{16J^2}{16J^2+U^2}\cos^2(\sqrt{16J^2+U^2}\frac{t_{bs}}{2})\nonumber
\\&&\qquad+\frac{U^2}{16J^2+U^2}.\nonumber
\end{eqnarray}
The optimal choice for the BS time is $t_{bs}=
\pi/\sqrt{16J^2+U^2}$ for which $q_{bs}=U^2/(16J^2+U^2)$. If the
hopping term is not controlled perfectly accurately but fluctuates
by $\delta J$ around a mean $J$ we set
$t_{bs}=\pi/\sqrt{16J^2+U^2}$ and obtain \equref{eq:q-bs}.


\begin{thebibliography}{00}
\bibitem{Foot1}
If the resonant state is either $\ket{aa}$ or $\ket{bb}$ then
there is also a suitable sequence of few pulses, which does not
require phase locking, interleaved with suitable loss intervals:
(i) a $\pi$ pulse that swaps $\ket{aa}$ and $\ket{bb}$ (ii) a
$\pi/2$ pulse that takes $\ket{ab}+\ket{ba}$ to
$\ket{aa}-\ket{bb}$, followed by (iii) a second $\pi$ pulse.
\bibitem{Foot2}
In a perfect experiment an even number $2j$ of atoms (two per
antisymmetric pair) is always detected but in the presence of
detector errord an odd number of atoms might arise.
\end{thebibliography}
\end{document}